\documentclass[twocolumn,showpacs,preprintnumbers]{revtex4}
\usepackage{amssymb}
\usepackage{graphicx}
\usepackage{dcolumn}
\usepackage{bm}

\begin{document}

\title{Electromagnetic properties of graphene junctions}
\author{S.~E.~Shafranjuk}
\homepage{http://kyiv.phys.northwestern.edu}
\affiliation{Department of Physics and Astronomy, Northwestern University, Evanston, IL 60208}
\date{\today }

\begin{abstract}
A resonant chiral tunneling (CT) across a graphene junction (GJ) induced by an external electromagnetic field (EF) is studied. Modulation of the electron and hole wavefunction phases $\varphi$ by the external EF during the CT processes strongly impacts the CT directional diagram. Therefore the a.c. transport characteristics of GJs depend on the EF polarization and frequency considerably.  The GJ  shows great promises for various nanoelectronic applications working in the THz diapason.
\end{abstract}

\maketitle

Unconventional electromagnetic properties of monoatomic layer graphene\cite{Novos,McCann} show
remarkable perspectives for scientific research and novel electronic applications. An external electromagnetic field (EF) affects the electric charge transport\cite{My-PRB} across the graphene junction (GJ) substantially. The mechanism of such a strong influence implies an intrinsic origin, since the external EF modulates the wavefunction phase of the charge carriers during their transmission across the GJ shown in Fig. 1(a). The charge carriers in graphene behave as relativistic massless 'chiral fermions' (CF) characterized by a 'pseudospin'%
\cite{Novos,McCann}. The pseudospin relates \cite{Novos,McCann} the CF particles to a certain sublattice A or B and is responsible for various unconventional d.c. electronic and magnetic properties of the monoatomic layer
graphene\cite{Novos2}. Significance of the pseudospin is proclaimed in the graphene junction (GJ) formed by a graphene sheet with two electrodes attached  as shown in the main Fig. 1. The electrodes can either be metallic or made of the graphene itself. The potential barrier is induced by the gate voltage $V_{\rm G}$ from a Si gate. The chiral tunneling (CT) strongly depends on the azimuthal angle $\phi$ between the two-dimensional electron momentum ${\bf p}=(\hbar k, \hbar q)$ and $\hat x$ axis (see Fig. 1). The CT is intrinsically related to a relativistic nature of CF since the electron and hole wavefunctions are interconnected in the monoatomic graphene\cite{Novos,McCann}. At some selected angles $\phi=\phi_n$ the CT probability becomes ideal ($T=1$). Such a resonant tunneling occurs due to a constructive quantum interference between an incoming electron ($e$) and a hole ($h$) moving inside the graphene barrier in a reverse direction and is characterized by the same pseudospin as sketched in Fig. 1 (Klein paradox\cite{Strange,Krekora}). The interference pattern resulted from the chiral tunneling (CT) is very sensitive to the phase difference $\varphi $ between $e$ and $h$ wavefunctions. Since an external electromagnetic field (EF) affects the phase $\varphi$ directly, a thorough study of that quantum interference mechanism between EF, $e$ and $h$ is necessary.
\begin{figure}
 \includegraphics{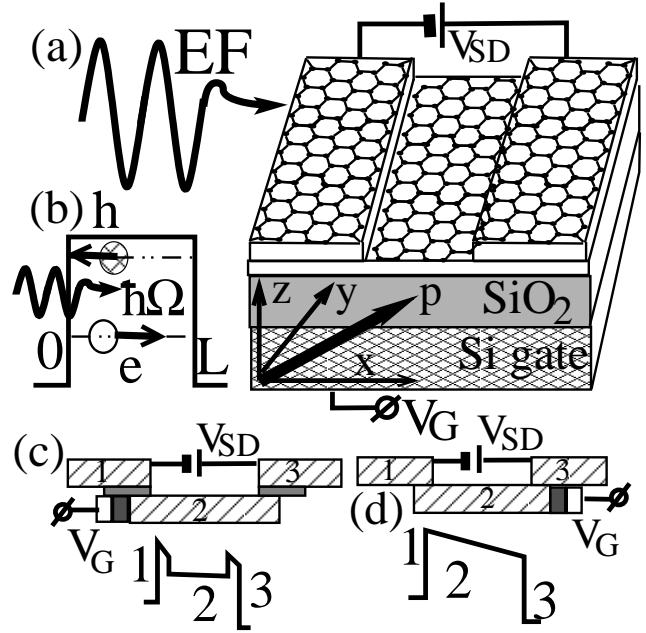}% Here is how to import EPS art
 \caption{\label{fig:Setup_b}
The graphene junction (GJ) controlled by the gate voltage $V_{\rm G}$ being exposed to the external electromagnetic field (EF). Two different barrier configurations (a)  with additional dielectric sublayers and (b) without them. Below we show the corresponding graphene barrier profiles.}
\end{figure}

This Letter addresses the EF-induced directional resonant chiral tunneling across the GJ. We expect that angular dependence of the differential tunneling conductivity $\sigma \left( \phi \right) $ ($\phi $ is the angle between the electric current $\mathbf{j}$ and the $x$-direction indicated in Fig. 1) is strongly influenced by the EF since the latter modulates the wavefunction $e-h$ phaseshift $\varphi $ inside GJ. In this way the a.c. field adjusts the resonant tunneling condition, causing a directional shift of the resonances. Initially we compute the steady state $\sigma \left( \phi ,V_{\rm SD}\right) $ curves for a GJ with either metallic or graphene electrodes biased by the source drain voltage $V_{\rm SD}$. The results obtained  for  the steady state  are then utilized for studying of the a.c. properties on the next stage. In particular we will see that the EF suppresses the CT current in the straightforward direction ($\phi =0$) due to an angular redistribution of the d.c. current. The d.c. current  increases to its resonant value at finite angles $\phi \neq 0$ instead. Below we focus on a weak a.c. field limit and discuss origin of the intrinsic noise. Here we examine two different setups of the voltage biased graphene barrier. One setup (quoted as the rectangular barrier) corresponds to a graphene barrier stripe 2 separated from the electrodes 1 and 3 [see Fig. 1(a)] by two thin dielectric sublayers at $x=0$ and $x=D$. Since resistivity of the dielectric sublayer is several orders of magnitude higher that of the graphene stripe, the bias voltage $V_{\rm SD}$ drops mostly on the formers [see the profile below Fig. 1(c)].  For the above reasons the d.c. electric field $\mathcal{E}$ is negligibly small inside the graphene section of the barrier. In this case the CF wavefunction $\hat{\Psi}$ inside the barrier graphene stripe is well approximated by plane waves. Another setup constitutes a bare monoatomic graphene stripe which contacts the attached graphene electrodes immediately as sketched in Fig. 1(d). The source drain bias voltage $V_{\rm SD}$ in this configuration drops entirely on the middle graphene stripe [see its profile below Fig. 1(d)], which corresponds to a trapezoidal shape of the graphene barrier tilted proportionally to $V_{\rm SD}$.   For the GJ having finite dimensions, the motion of CF is quantized. The quantization imposes additional constrains on the directional tunneling diagram. 

The Dirac equation for fermions is written as%
\begin{equation}
-i\hbar v\left( \left(\hat \sigma _{x}\otimes \hat 1\right) \partial _{x}+\left(\hat
\sigma _{y}\otimes \hat \tau _{z}\right) \partial _{y}\right) \Psi +eU\left(
x\right) \Psi =\varepsilon \Psi   \label{Dirac_A}
\end{equation}%
where $v=c/300$ is the massless fermion speed, $\varepsilon $ is the CF excitation energy, $\hat \sigma _{i}$ and $\hat \tau_k$ are the Pauli matrices, $\{i,k\}=1\dots3$, the barrier potential $U(x)$ is induced by the gate voltage $V_{\rm G}$. In the steady state, when the EF is off, the electric
current is fully suppressed when $V_{\rm SD}<U_{0}$ (for typical gate voltage $V_{\rm G}=1$ V and the SiO$_2$ thickness $d=300$ nm one finds\cite{Novos2} $U_0=2$ meV).
The CF excitation energy in the leads ($x<0$ and $x>L$) reads $\varepsilon_p =eV_{\infty }\pm \hbar v\sqrt{k^{2}+q_{n}^{2}}=\hbar v\left( k_{Fm}\pm \sqrt{%
k^{2}+q_{n}^{2}}\right) $. Setting $\varepsilon_p =0$ for a normal incidence ($%
q_{n}=0$) one writes $eV_{\infty }\pm \hbar vk_{Fm}=0$ or $eV_{\infty }=\hbar
vk_{Fm}$. If the leads are metallic, one writes $k=\sqrt{\left( \left( \varepsilon_p -\mu \right) /\hbar v\right) ^{2}-q_{n}^{2}}$. The energy inside the graphene stripe is $\varepsilon_p =eV_{\mathrm{G}}\pm \hbar v \sqrt{\tilde{k}^{2}+q_{n}^{2}}$. Inside the graphene barrier the CF wave vector is $\tilde{k}=\sqrt{\left( \left( \varepsilon_p -eU_{0}\right)/\hbar v\right) ^{2}-q_{n}^{2}}$. The transverse quantization conforms to the boundary conditions at $y=0$ and $y=W$. The transmission amplitude $t_{\varepsilon }$ is conveniently computed using a simple approach, which represents the CF wavefunctions $\hat \Psi $ as plane waves for a rectangular barrier and via Airy functions for a trapezoidal barrier. In particular, the scattering state for a trapezoidal barrier is constructed as 
\begin{eqnarray}
\Psi =\theta ( -x) [ \chi _{n,k}{\rm Ai}(k,x)+r_{n}\chi _{n,-k}{\rm Bi}(k,x)] \nonumber \\
+\theta ( x-L) t_{n}\chi _{n,k}{\rm Ai}(k, x-L)+\theta ( x) \theta ( L-x) \nonumber \\
\times [ \alpha _{n}\chi _{n,\tilde{k}}{\rm Ai}(\tilde{k}, x)e^{i\varphi } 
+\beta _{n}\chi _{n,-\tilde{k}}{\rm Bi}(\tilde{k},x)e^{i\varphi }]
\end{eqnarray}
where ${\rm Ai}(k,x)$ and ${\rm Bi}(k,x)$ are the Airy functions, and we also introduced the auxiliary functions $\chi _{n,k}\left( y\right) =a_{n}\left\vert \uparrow \right\rangle \otimes \left( \left\vert \uparrow \right\rangle +z_{n,k}\left\vert \downarrow \right\rangle \right) e^{iq_{n}y} $ $+a_{n}^{\prime }\left\vert \downarrow \right\rangle  \otimes \left( z_{n,k}\left\vert \uparrow \right\rangle +\left\vert \downarrow \right\rangle \right)  e^{iq_{n}y}$ $+b_{n}\left\vert \uparrow \right\rangle  \otimes \left( z_{n,k}\left\vert \uparrow \right\rangle
+\left\vert \downarrow \right\rangle \right) e^{-iq_{n}y}$ $+b_{n}^{\prime }\left\vert \downarrow \right\rangle \otimes \left( \left\vert \uparrow \right\rangle +z_{n,k}\left\vert \downarrow \right\rangle \right) e^{-iq_{n}y}$ where $\otimes $ is the Kronecker product, $\left\vert \uparrow \right\rangle ^{T}=\left( \begin{array}{cc}1 & 0
\end{array}\right) $ and $\left\vert \downarrow \right\rangle ^{T}=\left( \begin{array}{cc}
0 & 1 \end{array}\right) $ are 1$\times $2 matrices, $z_{n,k}=\pm \left( k+iq_{n}\right) /\sqrt{k^{2}+q_{n}^{2}}$, where $\pm $ signs apply to conductive (valence) bands, $k$ is positive for conductance band and negative for the valence band, the factor $z_{n,k}$ satisfies the identity $z_{n,k}z_{n,-k}=-1$. Permitted values of the angle $\tilde{\chi}$ inside the graphene stripe of a finite width $W$ are obtained from boundary conditions along the $y$-direction, $p_{y}=\tilde{q}_{n}=n\pi /W$, $W$ being the graphene stripe width. This gives the quantized angle $\tilde{\chi}_{n}$ inside the graphene stripe as $\tilde{\chi}_{n}=\arctan \left[n\pi /\left(k_{F}W\right)\right] $.  

The electric current between the electrodes 1  and 3  [see Fig. 1 (c,d)] is computed implementing methods\cite{Yeyati,Keldysh,Datta}. as $j=2\pi veN\left( 0\right) \sum_{p}\int d\varepsilon \left[ G_{3}^{K}\left( \varepsilon,{\bf p} \right) -G_{1}^{K}\left(
\varepsilon,{\bf p} \right) \right] $ where ${\bf p}=(\hbar k,\hbar q)$, $v=8.1\cdot 10^{5}$ m/s is the Fermi velocity in graphene, $N\left(0\right) =3.8$ nm$^{-1}$eV$^{-1}$ is the
electron density of states at the Fermi level, $G_{r}^{K}\left( \varepsilon,{\bf p}
\right) =-i\sum_{p}\left\vert t_{pp^{\prime }}\right\vert
^{2}e^{iq_n y}e^{ikD}\left( 2n_{p}-1\right) \delta \left( \varepsilon
-\varepsilon _{p}+\delta _{r,3}eV_{\rm SD}\right) $ is the Keldysh Green function%
\cite{Keldysh}, $r$ is the electrode index, $n_{p}$ is the distribution
function of electrons with a 2D momentum ${\bf p}=(\hbar k,\hbar q_n)$ assuming $q$ is quantized. Then one obtains $j=(\pi /2)evN\left( 0\right) \int d\varepsilon \sum_{p}\left\vert t_{\varepsilon
}\right\vert ^{2}(\left( 2n_{p}-1\right) \delta \left( \varepsilon
-\varepsilon _{p}+eV\right) $ $-\left( 2n_{p}-1\right) \delta \left(
\varepsilon -\varepsilon _{p}\right) )$ $=$ $\pi evN\left( 0\right) \sum_{n} \int
d\varepsilon \left\vert t_{\varepsilon ,n}\right\vert ^{2}\left(
n_{\varepsilon ,n}-n_{\varepsilon -eV,n}\right) $ where $N\left( 0\right)$ is the electron density of states at the Fermi level.
\begin{figure}
 \includegraphics{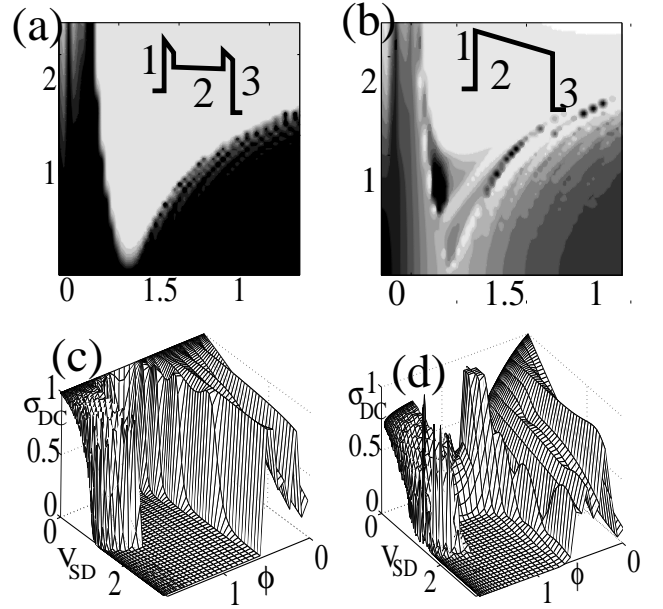}% Here is how to import EPS art
 \caption{\label{fig:dc_graphs}
The steady state transparency $\left\vert t\left( \varepsilon ,\phi \right) \right\vert ^{2}$ versus $0<\varepsilon <2.5$ and the incidence angle $0<\phi <\pi /2$ for the rectangular and trapezoidal barriers. The contour plot of $\sigma _{\rm SD} $ [in units of $e^2 v_F N\left( 0\right) $] versus $V_{\rm SD}$ (in units of the barrier height $eU_0$) (a) for a rectangular barrier,  and  (b)  for a trapezoidal barrier.}
\end{figure}
The steady state transparency diagrams $T = \left\vert t\left( \varepsilon ,\phi \right) \right\vert ^{2}$ versus $0<\varepsilon <2.5$ and the incidence angle $0<\phi <\pi /2$ for a rectangular and a trapezoidal barrier are shown in Fig. 2. In the contour plots Fig. 2(a,b) one may distinguish two types of the quantization. (a) A conventional quantization in the lateral direction  and  (b) a quantization taking place in the straightforward direction. The latter quantization is highly unconventional since it exists for the chiral fermions only. Corresponding scattering states are sharply angular-dependent, which is evident from the same contour plot. The 3D plots in Fig. 2(c,d) indicate numerous resonances of ideal transparency [dark areas and spots visible in Fig. 2(a,b) corresponding to $T=1$] occurring below the barrier when $V_{\rm SD}<2$ (in units of the barrier height $U_0$) at finite incidence angles $\phi \neq 0$.

If an external a.c. field is applied to the graphene junction, its influence strongly depends on the field intensity and polarization. The fermion wavefunction phase difference across the junction becomes modulated by the a.c. field $\psi \left( x,t\right) \rightarrow \psi \left( x,t\right) \exp \left( ieV_{\rm ac}\int^{t}dt\cos \Omega t\right) $. Below we discuss various a.c. field polarizations for the rectangular and trapezoidal barrier configurations. First we address the d.c. biased rectangular barrier shown in Fig. 1(c). If the electric field vector $\mathbf{E}$ of an external a.c. field is polarized as $\mathbf{E}=(E_{x},0,0)$, one may account for the a.c. field influence by setting $V_{\rm SD}\rightarrow V_{\rm SD}+eV_{ac}\cos \Omega \tau $, where $V_{ac}=E_{x}D$, $E_{x}=-\nabla _{x}\Phi -\dot{A}_{x}$, $\Phi $ and $A_{x}$ being the scalar and vector potentials. That approximation gives  $\hat{\Psi}\left( x,t\right) \rightarrow \hat{\Psi}\left( x,t\right) \cdot
\exp \{ieV_{ac}\int^{t}d\tau \cos \Omega \tau \}$. In this configuration the a.c. field does not
affect the barrier shape, it simply splits the quantized scattering resonant levels \cite{My-PRB}
only, i.e. $E_{0}\rightarrow E_{0}\pm n\Omega $ where $E_{0}$ is a
resonance energy, $n$ is the number of photons. That is the case
when EF modulates the quantum mechanical phase difference $\varphi $
across the junction. Using the equality\cite{Korn} $e^{i\alpha \sin \Omega t}\equiv \sum_{n}J_{n}\left( \alpha \right) e^{in\Omega t}$, one obtains the photon-assisted tunneling (PAT) differential
conductivity as $\sigma \left( V_{\rm SD}\right) =\sum_{n}J_{n}^{2}\left(
eV_{ac}/\hbar \Omega \right) \sigma _{0}\left( V_{\rm SD}-n\hbar \Omega
/e\right) $ where $\sigma _{0}\left( V_{\rm SD}\right) $ is the steady state
differential tunneling conductivity computed above, $V_{ac}$ is amplitude of
the a.c. bias voltage induced across the GJ by EF. The photon-assisted
tunneling is spectacularly proclaimed for the quantized scattering states. A
qualitatively different phenomena emerges if the external a.c. field applied 
to GJ modulates the barrier$^{\prime}$s height and shape. The barrier transparency $T$ in
that case is immediately affected by the a.c. field, polarized, e.g., as ${\bf E}=(0,0,E_z)$. For a weak field $%
V_{ac}<<U_{0}$, the PAT correction to the d.c. tunneling current is $%
j_{1}=2eN\left( 0\right) \int d\varepsilon \left\vert t_{\varepsilon ,\Omega
}^{\left( 1\right) }\right\vert ^{2}\left( 2n_{\varepsilon }-n_{\varepsilon
+\Omega -eV}-n_{\varepsilon -\Omega -eV}\right) $, where the angular-dependent transmission
amplitude $t_{\varepsilon ,\Omega }^{\left( 1\right) }$ is obtained from non-stationary boundary conditions at $x=0$ and $x=L$ for the two configurations of the  d.c. biased graphene barrier  shown in Figs 1(c,d). Since the EF modulates phase of the transmitted CFs, it  modifies also the condition of the resonance chiral tunneling. The a.c. field modulates the phase difference between the incident electron ($e$) and the reverse-propagating hole ($h$) waves propagating in opposite directions deflecting them inside the barrier  [as sketched in Fig. 3(d)]. This causes splitting of the CF momentum inside the barrier as $\hbar \tilde{k}_{\pm}=\pm \sqrt{((\varepsilon \pm \hbar \Omega -e U_0)/\hbar v)^2-\varepsilon ^2 \sin^2(\phi)}$ where typical frequencies are in the THz diapason. The wavefunction phaseshift inside the barrier is determined as $\tilde{\varphi}_{\pm}=\arctan \left[\varepsilon \sin \phi /(\hbar \tilde{k}_{\pm})\right] $ where $\phi $ is the electron incidence angle in the electrode attached to the graphene stripe. That means that the directional diagram strongly depends on the external field frequency $\Omega $ (typically 0.5-50 THz) and magnitude $V_{ac}$. In this way one gets a unique electromagnetic sensing devise, which tunning requires no energy excitation of the charge carriers, unlikely to most of the known sensors nowadays. 
\begin{figure}
 \includegraphics{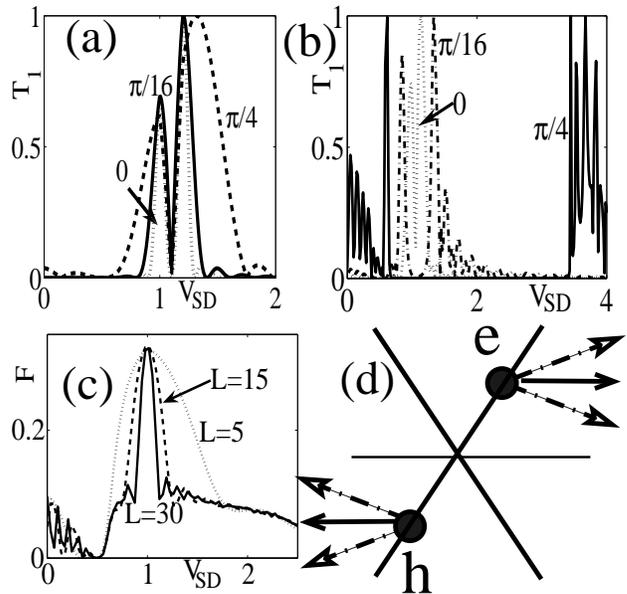}% Here is how to import EPS art
 \caption{\label{fig:AC_field} The EF-induced excess angular-dependent transparency $T_1$ of the rectangular (a) and trapezoidal (b) graphene barriers versus $V_{\rm SD}$. (c) Fano factor $F (V_{\rm SD})$ of the intrinsic Poisson noise for three different lengths $L=5, 15, 30$ of the trapezoidal barrier. (d) The EF-induced deflection of the electrons and holes during the chiral tunneling. Here $T_1$ and $F$ are expressed in dimensionless units while $V_{\rm SD}$ is in units of the barrier height $eU_0$.}
\end{figure}

The angular-dependent excess transmission $T_1 (\varepsilon ,\phi )$ induced by an a.c.  field,  which modulates the graphene barrier height $U_0$ is shown in Fig. 3 where the plots (a) and (b) correspond to the rectangular and trapezoidal barriers. The a.c. field tends to suppress tunneling in the longitudinal direction ($\phi =0$) while creates new resonances with $T=1$ at finite angles $\phi neq 0$.

Intrinsic noise in the monolayer graphene junction originates as follows. The thermal noise produced by phonons emitted in the electron-phonon collisions is depleted. The depletion is governed by quantum mechanical selection rules for the chirality conservation. That maintains the d.c. transport being
ballistic up to room temperatures. The intrinsic mechanism of the thermal noise depletion is as follows\cite{Ando}. Matrix element of the electron-phonon collisions according to Ref. \cite{Ando, McEuen} is $%
M_{pp^{\prime }}\propto \left\langle p\left\vert M\left( x\right) \right\vert p^{\prime }\right\rangle \cos \left( \phi _{pp^{\prime }}/2\right) $ where $\phi _{pp^{\prime }}$ is the angle between the initial and final states. The phase factor $\cos \left( \phi _{pp^{\prime }}/2\right) $ ensures suppression of the electron-phonon and electron-impurity collisions. Therefore the charge transport remains ballistic up to room temperatures. Another intrinsic noise (Poisson noise) arises due to the 'Zitterbewegung` effect caused by a jittering motion of the change carriers when electrons are randomly converted to holes forth and back. The intrinsic noise characteristic are illustrated by the Fano factor $F\left(V_{\rm SD}\right) $. The Fano factor is computed using the above expressions for the junction transmission coefficient $t_{pp^{\prime}}$ and the electric current. The noise persists even in the zero temperature limit. Such a Poisson noise is characterized by the Fano factor $F=\sum_{n}T_{n}\left( 1-T_{n}\right) /\sum_{n}T_{n}$, where $T_{n}$ is the transmission probability in the $n$-th channel and the summation is performed over all the conducting channels (in our setup this means just integration over $\phi $). From the plot $F(V_{\rm SD})$ shown in Fig. 3(c) for different length of the graphene barrier $L=5, 15, 30$ [in units of $v \hbar/(2 \pi e U_{0})$, which is typically 1-6 nm] one infers that the Poisson noise has a sharp resonance for long junctions ($L  > 30$) and substantially decreases when $V_{\rm SD}\geq U_{0}$. 

In conclusion we computed the electric current across a monoatomic layer graphene junction in conditions when an external electromagnetic field is applied. We find that the a.c. properties of GJ strongly depend on the EF frequency and amplitude. The EF induces an angular redistribution of the tunneling current. Such a deflection  comes rather from a quantum mechanical phase shift than from an inelastic excitation by the EF. That ensures no heating is involved during the absorption. The a.c. current induced by EF across the GJ has a sharp angular dependence, which potentially can be exploited in sensor nanodevices of the external electromagnetic field. An experimental observation of such a spectacular EF- induced  deflection would provide strong supporting evidences for the chiral fermion concept in the monoatomic graphene. The EF-induced CT resonances improve the device switching speed and enhance the frequency resolution. The phenomena considered above have a great potential for building of fast-switching transistors and EF sensors working in the THz diapason. 

I wish to thank V. Chandrasekhar and P. Barbara for fruitful discussions.
This work had been supported by the AFOSR grant FA9550-06-1-0366.


\begin{thebibliography}{99}
\bibitem{Novos} K. S. Novoselov, E. McCann, S. Morozov, V. I. Falko, M. I.
Katsnelson, U. Zeitler, D. Jiang, F. Schedin, and A. K. Geim, Nature Phys. 
\textbf{2}, 177 (2006).

\bibitem{McCann} E. McCann and V. I. Falko, Phys. Rev. Lett. \textbf{96},
086805 (2006).

\bibitem{My-PRB} S. E. Shafranjuk, Phys. Rev. B\textbf{76}, 085317 (2007).

\bibitem{Strange} P. Strange, Relativistic Quantum Mechanics (Cambridge University Press, Cambridge, UK, 1998). 

\bibitem{Krekora}P. Krekora, Q. Su and R. Grobe, Phys. Rev. Lett. {\bf 92}, 040406 (2004).

\bibitem{Novos2} K. S. Novoselov, A. K. Geim, S. V. Morozov, D. Jiang, Y.
Zhang, S. V. Dubonos, I. V. Grigorieva, A. A. Firsov, Science \textbf{306},
666 (2004);

\bibitem{Yeyati} A. L. Yeyati and M. B\"uttiker, Phys. Rev. B \textbf{52},
R14360 (1995).

\bibitem{Keldysh} L. V. Keldysh, Sov. Phys. JETP \textbf{20}, 1018 (1965).

\bibitem{Datta} S. Datta, Electronic Transport in Mesoscopic Systems
(Cambridge University Press, Cambridge, UK, 1997).

\bibitem{Kats-C} M. I. Katsnelson, K. S. Novoselov, and A. K. Geim, Nature
Phys. \textbf{2}, 620 (2006).

\bibitem{Korn} G. A. Korn and T. M. Korn, Handbook for Scientists and
Engineers, McGraw-Hill (1967).

\bibitem{Ando} T. Ando, T. Nakanishi, and R. Saito, J. Phys. Soc. Jpn., 
\textbf{67}, 2857 (1998); T. Ando, T. Nakanishi, J. Phys. Soc. Jpn., \textbf{%
67}, 1704 (1998).

\bibitem{McEuen} P. L. McEuen, M. Bockrath, D. H. Cobden, Y. G. Yoon, and S.
Louie, Phys. Rev. Lett., \textbf{83}, 5098 (1999).
\end{thebibliography}
\end{document}